# Modelling and design of highly coupled piezoelectric energy harvesters for broadband applications

D Gibus[1,2], P Gasnier[1], A Morel[1,2], S Boisseau[1] and A Badel[2]

[1]Université Grenoble Alpes, CEA, LETI, MINATEC, F-38000 Grenoble, FRANCE
[2]Université Savoie Mont Blanc, SYMME, F-74000 Annecy, FRANCE

david.gibus@cea.fr

**Abstract.** This paper reports a method to design highly coupled piezoelectric energy harvesters with frequency tuning capabilities using nonlinear electrical techniques. A cantilever beam with two PMN-PT patches has been optimized thanks to both analytical modelling and Finite Element Methods (FEM). The built prototype exhibits a strong electromechanical coupling ($k^2$=17.6%) and a figure of merit ($k_m^2Q$=12.4) which allow a bandwidth corresponding to 22% of the resonant frequency value.

## 1. Introduction

Ambient energy harvesting appears as a relevant solution to supply electrical energy to sensors where batteries cannot be used or need to be complemented. In environments without light or thermal gradient, vibration energy harvesting is necessary and piezoelectric systems based on mechanical resonators are interesting for their high power density at small scale [1]. However, the frequency bandwidth limits are still an important issue.

Electrical nonlinear methods have recently been proposed to enlarge the harvesting bandwidth by tuning the resonant frequency of linear piezoelectric harvesters [2,3]. In opposition to nonlinear mechanical harvesters [4], these techniques are not dependent on the input excitation level and are based on the influence between the mechanical resonator dynamics and the electrical circuit. For that purpose, the use of highly-coupled piezoelectric generators was shown to be mandatory to enlarge the frequency response. However, few works in prior art have studied such generators, since increasing the electromechanical coupling ($k^2$) beyond a certain value did not show any interest to improve the maximal harvested power [1], and because the bandwidth was rarely studied.

In 2013, Ahmed-Seddik *et al.* [5] proposed a capacitance tuning technique and a 33-mode PZN-PT prototype ($k^2$=49.3%). The structure has though a low quality factor and may not be adapted to industrial process as it seems complicated to assemble. In 2014, Badel and Lefeuvre [2] explained the interest of maximizing the electromechanical coefficient to improve the bandwidth thanks to a non-linear electrical technique. They also introduced a strongly coupled ($k^2$=53%) cantilever-based PZN-PT harvester, but no design method was proposed. Hence, we propose a design method based on both analytical modelling and Finite Element Methods (FEM) to build highly-coupled vibration energy harvesters. The built prototype reaches a high value of the conventional figure of merit for such harvesters $k_m^2Q$.

## 2. Model

Like [2], our device implements a long tip mass (figure 1). Nevertheless, the single degree of freedom (SDOF) model is inaccurate for this particular geometry [6] and the commonly used distributed parameter model [7] is inappropriate for optimization and needs numerical resolutions to compute the resonant frequencies and deduce the coupling. Thus, we propose a 2-degree-of-freedom (2-DOF) model to design highly coupled bimorph cantilevers with a long tip mass.

The proposed model, which is an evolution of the one presented by [6], is based on the Euler-Bernoulli assumptions and the mass of the beam is neglected compared to the mass of the proof mass. In addition to the analysis of the deflection $u$ and the force $F$ performed in the SDOF model, the effect of the rotation $\theta_{Lb}$ of the tip mass and the resultant couple $C$ are analysed (figure 2). Both tip mass rotary inertia $I_t$ and the distance between its center of gravity and the end of the beam are taken into account [6]. Moreover, electrodes cover piezoelectric patches at the bottom and the top surfaces. They are connected in parallel and the electric field is not considered uniform along the piezoelectric thickness as presented in [8].

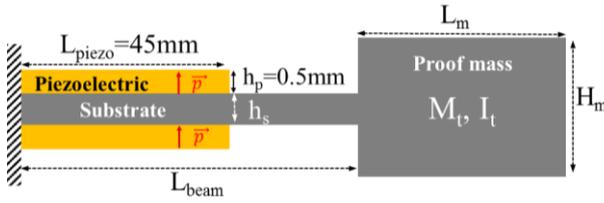 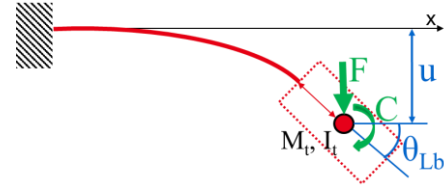

**Figure 1.** Parameters of the bimorph cantilever.       **Figure 2.** Cantilever during bending.

The constitutive equations of piezoelectric and substrate materials are used to express the internal bending moment of the beam according to the plane stress or the plane strain assumption [8]. The beam stiffness matrix is then determined and the following system of electromechanical equations (1) is derived thanks to the force and torque equilibrium analyses and from the Kirchhoff's law:

$$\begin{bmatrix} M_t & 0 \\ 0 & I_t \end{bmatrix} \begin{pmatrix} \ddot{u} \\ \ddot{\theta}_{L_b} \end{pmatrix} + \begin{bmatrix} K_1 & K_2 \\ K_3 & K_4 \end{bmatrix} \begin{pmatrix} u \\ \theta_{L_b} \end{pmatrix} + \begin{bmatrix} \alpha_1 \\ \alpha_2 \end{bmatrix} V = \begin{bmatrix} M_t \gamma \\ 0 \end{bmatrix}$$
$$\alpha \dot{\theta}_{L_p} + C_p \dot{V} = I$$
(1)

where $\gamma$ is the excitation acceleration of the clamped end, $V$ and $I$ are the voltage across and the current through the electrodes respectively. $C_p$, $\alpha$, $\alpha_1$, $\alpha_2$ and the coefficients of the stiffness matrix $[K]$ depend on the geometry and the material coefficients. $M_t$ and $I_t$ are the mass and rotary inertia of the tip mass respectively. $\theta_{Lp}$ is the bending angle at the end of the piezoelectric patches ($x=L_{piezo}$).

The open and closed circuit resonant frequencies ($f_{oc}$ and $f_{sc}$) of the first vibration mode can be analytically expressed from (1) and thus, the corresponding global coupling coefficient can be calculated from (2) [9].

$$k^2 = \frac{f_{oc}^2 - f_{sc}^2}{f_{oc}^2}$$
(2)

## 3. Optimization for a given patch

Our model is used to find the geometrical parameters (figure 1) that maximize the first vibration mode coupling coefficient for given PMN-PT patches (45×10×0.5mm). The analytical expression of the global coupling simplifies the parametric study in comparison with the distributed parameters model or FEM simulations. The first resonant frequency was sought around 30Hz and the total length ($L_m+L_{beam}$) had to be less than 90mm. The beam and the mass are made of steel and the piezoelectric parameters are given in table 1.

**Table 1.** Material parameters (PMN-PT X2B of TRS ceramics). Material coupling coefficients $k_{31}^2$ are calculated from [9].

| $d_{31}$ (pm.V$^{-1}$) | $s_{11}^E$ (×10$^{-12}$Pa$^{-1}$) | $\epsilon_{33}^T$ (F.m$^{-1}$) | $k_{31}^{l\,2}$ plane stress | $k_{31}^{w\,2}$ plane strain |
|---|---|---|---|---|
| -699 | 52.1 | 5400 $\epsilon_0$ | 19.62% | 68.09% |

According to a parametric study using the 2-DOF model presented in section 2, the piezoelectric patches should be as long as the substrate beam. It also shows that increasing the mass length homogenizes the strain distribution, which explains why such geometries are advantageous. As depicted in figure 4, different optimal substrate thicknesses and coupling coefficients are found for plane stress and plane strain assumptions. Even if the proposed 2-DOF model fits well with 2D FEM simulations (Comsol), it is shown that two-dimensional analysis is not sufficient to predict the actual electromechanical coupling [10]. Indeed, there is a notable impact of the length/width ratio for strongly coupled cantilevers and a 3D FEM analysis is compulsory to design such generators. As this ratio is large for the PMN-PT patches, our model with plane stress hypothesis is closer to the 3D study.

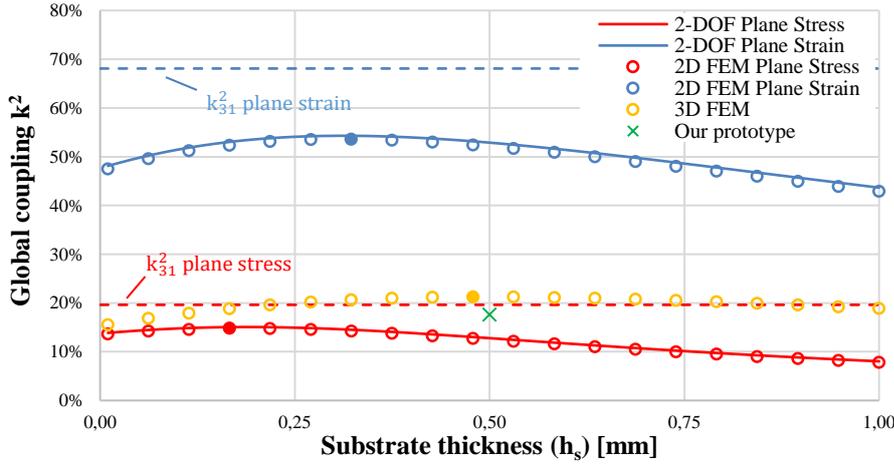

**Figure 3.** Global coupling coefficient vs substrate thickness for $H_m$=5mm, $L_{beam}$=$L_m$=45mm and beam width = 10mm for 3D FEM. Full dots are the maximal values.

## 4. Experiment results

The determined optimal substrate thickness is equal to 0.5mm and a prototype has been assembled (figure 4). Its admittance response has been measured thanks to an impedance analyser and has been matched with the model presented in [2] to get the coupling coefficient and the quality factor (table 2 and figure 5). The differences between the simulations and the experiments are probably due to the imperfectly clamped end and potential errors on material properties.

**Table 2.** Analytical, numerical and experimental results for the cantilever with the optimal parameters.

| | Short circuit resonance frequency | Open circuit resonance frequency | Global coupling coefficient $k^2$ | Modified coupling coefficient $k_m^2 = k^2/(1-k^2)$ | Mechanical quality factor $Q$ | Max power @ 0,08g |
|---|---|---|---|---|---|---|
| **2-DOF Plane Stress** | 31.46 Hz | 33.69 Hz | 12.78% | 14.65% | | |
| **2-DOF Plane Strain** | 36.74 Hz | 53.52 Hz | 52.89% | 112.26% | | |
| **3D FEM** | 34.32 Hz | 38.70 Hz | 21.37% | 27.17% | | |
| **Experimental results** | 28.61 Hz | 31.52 Hz | 17.61% | 21.37% | 58 | 370µW |

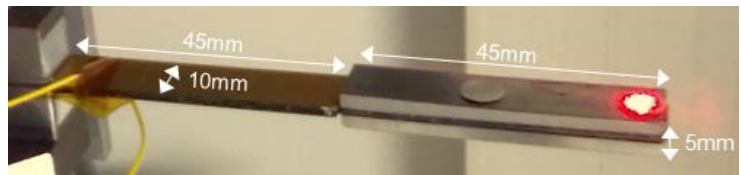

**Figure 4.** Fabricated prototype under test ($h_s$=$h_p$=0.5mm)

The product $k_m^2Q$ (computed from Table 2) reaches 12.4 for our prototype. It is much larger than what was obtained in a previously reported, more sophisticated, PMN-PT structure [6] ($k_m^2Q$=1.39). The cantilever presented in [2] exhibits a $k_m^2Q$ of 31.6 thanks to the better electromechanical properties of the PZN-PT compared to the PMN-PT we used.

The prototype has been excited under a constant acceleration of 0.02g between 25Hz and 38Hz. The root mean square output voltage of the device has been measured for 50 resistive loads (1kΩ to 35MΩ) and 50 frequency values in this band. As shown in figure 6, the harvester provides a maximal power of 32.7µW and a bandwidth equal to 8.4% of the resonant frequency by tuning the resistive load.

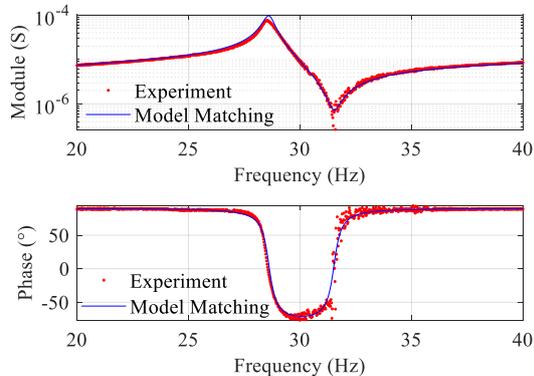
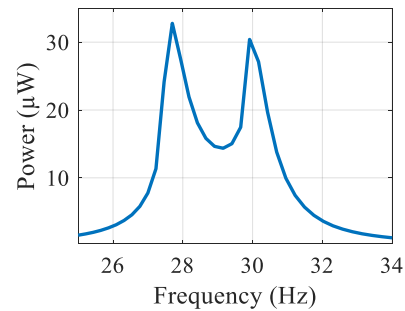

**Figure 5.** Measured admittance of the prototype and comparison with the model presented in [2].

**Figure 6.** Measured power vs frequency at optimal resistive load under 0.02g acceleration.

**Conclusion**
A fully analytical model has been developed to optimize the global electromechanical coupling of a vibration energy harvester. This highly coupled device is based on a cantilever beam with long tip proof mass and two PMN-PT patches. We discussed the validity of our model when the structures are highly coupled and the need to take the effect of the beam width into account. In this case, we emphasise that 3D simulations must be used to refine the design of the harvester. Our prototype exhibits a coupling of $k²=17\%$ and a figure of merit $k_m²Q=12.4$. With these characteristics, we expect a 22% frequency bandwidth when associated with a non-linear interface circuit such as the FTSECE [2] or the SC-SECE [3].